\documentclass[letterpaper,twocolumn,prl,aps,showpacs,superscriptaddress]{revtex4-1}
\usepackage[latin1]{inputenc}
\usepackage{bm}
\usepackage[usenames]{color}
\usepackage{multirow}
\usepackage{amssymb}
\usepackage{amsbsy}
\usepackage{amsmath}
\usepackage{stmaryrd}
\usepackage{graphicx}
\usepackage{epsfig}
\usepackage{placeins}
\makeatletter
\begin{document}

\title{Coherent Spin-Current Oscillations in Transverse Magnetic Fields}

\author{Robin Steinigeweg}
\email{r.steinigeweg@tu-bs.de}
\affiliation{Institute for Theoretical Physics, Technical University
Braunschweig, D-38106 Braunschweig, Germany}

\author{Stephan Langer}
\author{Fabian Heidrich-Meisner}
\affiliation{Department of Physics and Arnold Sommerfeld Center for Theoretical Physics, Ludwig-Maximilians-Universit\"at M\"unchen, D-80333 M\"unchen, Germany}

\author{Ian P.~McCulloch}
\affiliation{School of Physical Sciences, The University of Queensland,
Brisbane, QLD 4072, Australia}

\author{Wolfram Brenig}
\affiliation{Institute for Theoretical Physics, Technical University
Braunschweig, D-38106 Braunschweig, Germany}

\date{\today}

\begin{abstract}
We address the coherence of the dynamics of spin-currents with components
transverse to an external magnetic field for the spin-$1/2$ Heisenberg chain. We
study current autocorrelations at finite temperatures and the real-time dynamics
of currents at zero temperature. Besides a coherent Larmor oscillation, we find
an additional collective oscillation at higher frequencies, emerging as a
coherent many-magnon effect at low temperatures. Using numerical and analytical
methods, we analyze the oscillation frequency and decay time of this
coherent current-mode versus temperature and magnetic field.
\end{abstract}


\maketitle

\begin{figure}[tb]
\includegraphics[width=0.8\columnwidth]{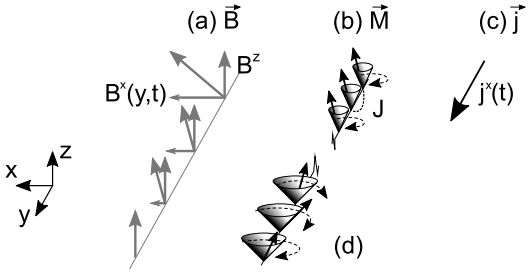}
\caption{
Quasi-classical sketch of transverse spin transport in a spin chain directed
along the $y$-direction: (a) External magnetic field with a static
`bias' component $B^z$ and a perturbing, space- and time-dependent component
$B^x(y,t)$; (b) $B^z$ produces a magnetization $\vec{M}$ with a \emph{homogeneous}
transverse component and a Larmor precession, $B^x(y,t)$ produces an
\emph{inhomogeneous} transverse component driving (c) a transverse spin-current
$j^x(y,t)$ in $y$-direction. The dynamics of $j^x(y,t)$ is controlled by the
intrinsic exchange coupling $J$. (d) As a function of time, the magnetization relaxes
and the transverse component dephases.} \label{setup}
\end{figure}

Controlling quantum coherence is paramount for future information processing
\cite{Fischer2009}. The coherence of localized quantum spin degrees of freedom
has been studied in a wide variety of systems, including semiconductor quantum
dots \cite{Reilly2008, Foletti2009}, molecular magnets \cite{Ardavan2007},
nitrogen vacancies in diamond \cite{Balasubramanian2009}, carbon nanotubes
\cite{Churchill2009}, and ultracold atoms \cite{Beugnon2007}. Coherence in spin
transport has been addressed primarily in semiconductors \cite{Kato2004}.
A new route into coherent spin transport may arise from
quantum magnets. Here, magnetization is transported solely by virtue of exchange
interactions and (de)coherence of spins will emerge as a purely intrinsic
many-body phenomenon. In one-dimensional (1D) spin systems, magnetic transport
has experienced an upsurge of interest in the last decade due to the discovery
of very large magnetic {\em heat} conduction \cite{sologubenko00}
and long nuclear magnetic relaxation times \cite{thurber01}. Genuine
{\em spin} transport in such materials remains yet to be observed experimentally
and if combined with materials with small exchange couplings \cite{klanjsek08,
kuehne10}, the coherent manipulation of spin transport using laboratory
magnetic fields may become feasible. Theoretically, spin transport has already
been given significant attention (see Refs.~\onlinecite{zotos-review, hm07a}
for a review), previous studies, however, have focused on the
longitudinal spin conductivities only, excluding the physics of
coherence. Therefore, in this Letter we investigate the dynamics of
spin-currents with components transverse to an externally applied magnetic
field, as sketched in Fig.~\ref{setup}. This setup allows us to study the
collective precession frequency of the transverse spin-current and its decay
time, which will be at the prime focus of this Letter. We will show that,
besides a coherent oscillation at the Larmor frequency, a second nontrivial
collective oscillation at higher frequencies emerges at low temperatures. This
oscillation is identified as a pure many-magnon effect and also becomes coherent
in the low temperature limit.

In this Letter, we study the antiferromagnetic Heisenberg spin chain,
one of the fundamental models to describe magnetic properties of interacting
electrons. It is relevant to the physics of low-dimensional quantum magnets
\cite{johnston00}, ultra-cold atoms \cite{trotzky08}, nanostructures
\cite{gambardella06}, and - seemingly unrelated - fields such as string
theory \cite{kruczenski04} and quantum Hall systems \cite{kim96}. The Hamiltonian reads

\begin{equation}
H = J \sum^N_{r,\alpha} S^\alpha_r S^\alpha_{r+1} - B^z \sum^N_r S^z_r \, .
\label{H}
\end{equation}
$S^\alpha_r$ ($\alpha = x, y, z$) are the components of spin-$1/2$ operators at site $r$, $N$
is the number of sites, $J>0$ is the exchange coupling constant, $B^z$ is
a longitudinal magnetic field, 
and $\hbar \equiv 1$ \cite{epaps}. For $B^z < B^z_c = 2 J$,
Eq.~(\ref{H}) implies a gapless Luttinger liquid \cite{Griffiths1964, Haldane1980}
and, for $B^z > B^z_c$, a gapped ferromagnetic ground state.

We investigate the transverse spin-current dynamics for two complementary scenarios
and use methods appropriate for each situation. First, we study current
autocorrelations at finite temperature, using numerically exact diagonalization
(ED) and an asymptotic analytic analysis (AAA). Second, applying the adaptive
time-dependent density matrix renormalization group (tDMRG) \cite{tDMRG}, we analyze
the real-time dynamics of currents at zero temperature during the evolution from
initial states with an inhomogeneous magnetization. Qualitatively, the same physics
explains our observations in both scenarios, and even a quantitative agreement can
be obtained.

We begin by discussing the current autocorrelations
$
\tilde{C}^{\mu\nu}(\omega) = \sum_{lm}
e^{-\beta E_m}
\langle l |j^\mu | m \rangle
\langle m |j^\nu | l \rangle
\delta(\omega - E_m + E_l)
/ Z N
$
\cite{Mahan}, where $| l \rangle$, $| m \rangle$ and $E_{l,m}$ are eigenstates
and -energies of Eq.~(\ref{H}), $\beta=1/T$ is the inverse temperature, $j^\mu$
is the zero-momentum spin current, $\mu, \nu = x, y, z$, $Z$ is the
partition function, and $\omega$ is the frequency. More precisely, we consider a
symmetrized version $C^{\mu\nu}(\omega) = [\tilde{C}^{\mu\nu}(\omega) +
\tilde{C}^{\mu\nu}(-\omega)]/2$, i.e., in the time domain, we focus on the real
part $C^{\mu\nu}(t) = \mathrm{Re}[\tilde{C}^{\mu\nu}(t)]$. While $B^z$ breaks
total spin conservation, a spin-current can still be defined by decomposing the
time derivative of the spin-density at momentum $q$ as $\partial_t S_q^\mu =
\partial_t S_q^\mu |_{J=0} - \imath q j^\mu_q$ into a local source (sink) term
due to $B^z$ (present without any exchange interactions) and the actual exchange
mediated spin-current $j^\mu_q$. The latter then derives from the continuity
equation for $S_q^\mu$ at $B^z = 0$. In turn $\vec{j} = \imath \sum_r \vec{S}_r
\times \vec{S}_{r+1}$, where $\vec{j} = \vec{j}_{q=0}$. The eigenstates
(energies) are classified according to the total spin $z$-component $\sum_r
S^z_r | l \rangle = M | l \rangle$. Since $| l \rangle$ is {\em independent} of
$B^z$ and $C^{\mu\nu}$ is diagonal at $B^z=0$, it will remain diagonal at any
$B^z$. Moreover, by symmetry $C^{xx} = C^{yy}$. However, since $j^x$ mediates
transitions between sectors with $\Delta M = \pm 1$ while $j^z$ conserves $M$,
the autocorrelations $C^{xx}$ (transverse) and $C^{zz}$ (longitudinal) {\em
differ} at $B^z \neq 0$. This difference is solely due to the field dependence
of the eigenenergies. Formally speaking, this aspect is at the center of this
Letter. For the remainder we abbreviate $C^{xx(zz)}$ by $C^{x(z)}$. 
Note that by the continuity equation the current autocorrelations at small $q$ are related
to the dynamic spin structure factor $S^{\mu\nu}(q,\omega)$, exhibiting a
matrix symmetry identical to $C^{\mu\nu}(\omega)$ at $B^z \neq 0$
\cite{Grossjohann2009}.

Generically, the longitudinal autocorrelation decomposes into a Drude weight $D^z$
and a regular part, i.e., $C^z(\omega) = D^z \delta(\omega) +
C^z_{\mathrm{reg}}(\omega)$. A significant body of evidence in favor of
$D^z(T \geq 0) \neq 0$ for $B^z < B^z_c$ has been gathered
\cite{zotos-review,hm07a,Sirker2009}, with thermally activated behavior of
$D^z(T)$ for $B^z > B^z_c$ \cite{hm07a}. Less is known on the specific shape of the regular
part \cite{Sirker2009}. The previous discussion of symmetries of $C^{\mu\nu}$ implies that
$C^x(\omega)= \sum_\pm [D^x \delta(\omega \pm B^z)+ C^x_{\mathrm{reg}}(\omega
\pm B^z)]$. In general, $D^x$ and $C^x_{\mathrm{reg}}$ will {\em not} be
identical to $D^z/2$ and $C^z_{\mathrm{reg}}/2$, respectively,
due to the different $B^z$-dependence of Boltzmann weights in $C^z$ and $C^x$.
In the time domain, $D^x$ implies a coherent, nondecaying
oscillation of the transverse current at the Larmor frequency $\omega_L = B^z$,
permitted by the integrability of
Eq.~(\ref{H})
\cite{zotos-review}. On the other hand, $C^x_{\mathrm{reg}}$ {\em a priory}
implies only decoherence and damping. In the following, however, we demonstrate
that at sufficiently
low $T$ and finite $B^z$, out of $C^x$, a new collective quasi-coherent oscillation
of the
current emerges.
The oscillation frequency differs from the `simple' Larmor frequency
and cannot be understood within a one-magnon picture.

\begin{figure}[tb]
\includegraphics[width=0.9\linewidth]{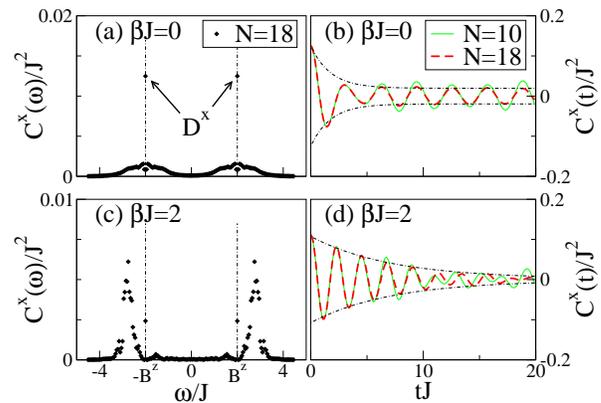}
\caption{(color online) Frequency and time dependence of the autocorrelation
$C^x$ for $B^z/J = 2$ and (a), (b) $\beta = 0$, (c), (d) $\beta J= 2$ (ED). The
Drude weight in (a), (c) is visible exactly at the Larmor frequency $\omega_L = B^z$
(vertical dashed-dotted lines). In (b), (d) the envelopes of fits [as defined in
Eq.~(\ref{eq:cxt})] to data for $N = 18$ are shown (dashed-dotted curves).}
\label{figure_Cx}
\end{figure}

First, we discuss high temperatures, i.e., $\beta = 0$. A straightforward
analysis yields
\begin{equation}
C^x(\omega) = [C^z(\omega-B^z) + C^z(\omega+B^z)]/2 \, . \label{Cx_beta0}
\end{equation}
Figure \ref{figure_Cx}~(a) displays $C^x(\omega)$ for $B^z/J = 2$. As can
be seen from Fig.~\ref{figure_Cx}~(b), this approximately transforms into $C^x(t)
\approx [R(t) + D^x] \cos(\omega_L t)$ in the time domain, with $R(t)$ {\em
rapidly} decaying within $\sim1/\omega_L$ and a `trivial' coherent oscillation
due to the Drude weight.

Next we reduce the temperature to $\beta J = 2$ at $B^z/J = 2$.
Figure \ref{figure_Cx}~(c) clearly
shows two effects. First, the Drude
weight $D^x$ is strongly reduced. This reduction continues monotonously with
increasing $B^z$ (as discussed within the AAA below). Second, the regular part
$C^x_{\mathrm{reg}}$ is strongly enhanced and undergoes an asymmetric weight
redistribution with a major peak developing at a frequency $\omega$ {\em larger}
than $\omega_L$ (and a minor peak at $\omega < \omega_L$).
This is consistent with $B^z$ breaking particle-hole symmetry. In the time
domain, see Fig.~\ref{figure_Cx}~(d), we find that
\begin{equation}
C^x(t) \approx R(t) \cos[(\omega_L + \delta \omega) t] + D^x \cos(\omega_L t)
\label{eq:cxt}
\end{equation}
allows for a reasonable leading-order fit of $C^x(t)$ over several oscillation
periods by assuming an exponential behavior $R(t)= R_0 \exp(-t/\tau)$, i.e., a
single decay time $\tau$. In fact, $C^x(\omega)$ is rather close to a Lorentzian
in Fig.~\ref{figure_Cx}~(c). We find this approach to apply at least to $B^z/J
\leq 3$ and $\beta J \leq 3$ and to have very little finite-size effects for
the system sizes ($N=10, \ldots, 18$) and time scales ($t J \leq 20$) studied
\cite{epaps}. Figure \ref{figure_Cx}~(c) is a central result of this Letter. It
unveils the emergence of a new collective frequency scale, namely at $\omega_L +
\delta \omega$, in the transverse transport process besides the Larmor frequency.
Moreover, for $\tau \rightarrow \infty$ this process would be coherent.

\begin{figure}[tb]
\includegraphics[width=0.8\linewidth]{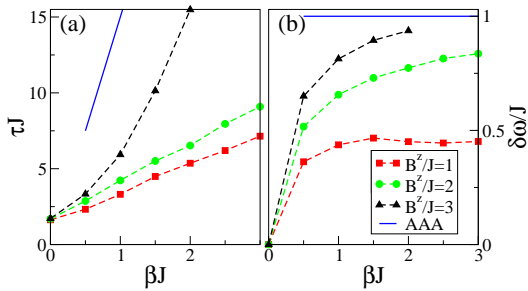}
\caption{(color online) (a) Decay time $\tau$ and (b) frequency shift
$\delta \omega$ w.r.t.~$\beta$ for different $B^z$. Data are extracted from the
autocorrelation $C^x(t)$, using ED for $N = 18$ (symbols). Solid lines represent
the low-temperature asymptote above the critical field, dashed curves are guides
to the eye.}
\label{figure_frequency_shift_decay_time}
\end{figure}

In Fig.~\ref{figure_frequency_shift_decay_time} we summarize our findings for
$\tau$ and $\delta \omega$ over a range of temperatures and fields of
$0 \leq \beta J \leq 3$ and $1 \leq B^z/J \leq 3$. Figure
\ref{figure_frequency_shift_decay_time}~(a) shows that $\tau$ increases roughly
linearly with $\beta$, with an increasing slope as $B^z$ increases. While finite
system studies will not clarify if this result implies true coherence for a
particular range of $B^z$ as $T \rightarrow 0$,
Fig.~\ref{figure_frequency_shift_decay_time}~(a) is at least strongly indicative
of a large
$\tau$ in that limit. Regarding
$\delta \omega$, Fig.~\ref{figure_frequency_shift_decay_time}~(b) clearly signals a
saturation roughly at $\beta J \sim B^z/J$ with the value at saturation
increasing with $B^z$. We emphasize that $\delta \omega \neq 0$ directly implies
that the transverse current {\em cannot} be described in terms of transitions between
the zero- and one-magnon sector.
The dominant spectral weight of such
one-magnon excitations at $q=0$ is exactly at $\omega_L$ \cite{Grossjohann2009},
leading to $\delta \omega=0$.

To gain insight into the origin of $\delta \omega \neq 0$ we present an
AAA for $B^z > B^z_c$ and low $T$. Here, the contribution of different $M$-sectors
to $C^x$ can be dissected asymptotically
and, after an extensive analysis \cite{epaps}, we arrive at a simple picture: the
transverse
current is carried dominantly by transitions from the one-magnon sector
around $q \sim \pi$ into antibound states of the \emph{two-magnon} continuum at
the same $q$.
The related frequencies $\sim \omega_L + J$ and also the asymptotic form of
$C^x(t)$ can be obtained analytically:
\begin{equation}
C^x_{\mathrm{AAA}}(t) \approx \sqrt{\frac{J^3}{\pi}} \, e^{-\beta(B^z -
2J)} \, \mathrm{Re} \left ( \frac{e^{\imath (\omega_L + J) t}}{\sqrt{2 \beta +
\imath t}} \right )
\, .
\label{asymp}
\end{equation}
This is consistent with Fig.~\ref{figure_frequency_shift_decay_time}~(b), which also
suggests that $\delta \omega \rightarrow J$ as $B^z$ increases and $\beta
\rightarrow \infty$. The thermal activation in Eq.~(\ref{asymp}) stems from the
one-magnon energy at $q = \pi$. The damping results from summing over all
transitions in the vicinity of $q = \pi$ and its power-law behavior, i.e.~$\sim
t^{-1/2}$, for $t \rightarrow \infty$ clearly shows that the single-scale
exponential used for $R(t)$ in Eq.~(\ref{eq:cxt}) is an approximation for not
too low
temperatures
$T$ only. Nevertheless, for a comparison with
Fig.~\ref{figure_frequency_shift_decay_time}~(a), we extract a `decay time' from
the envelope of Eq.~(\ref{asymp}), i.e.~$| C^x_{\mathrm{asy}}(t) /
C^x_{\mathrm{asy}}(0)| \leq 1/e$ for $t \geq \tau$, leading to $\tau = 2 \beta
\sqrt{e^4 - 1} \approx 15 \beta$.
As shown in Fig.~\ref{figure_frequency_shift_decay_time}~(a) (straight line),
our ED data for $B^z/J = 3$ are consistent with the asymptotic line, e.g.,
the slope $\mathrm{d}\tau / \mathrm{d}\beta$ is already close to $15 \beta$ at
$\beta J \sim 2$.

\begin{figure}[tb]
\includegraphics[width=0.9\linewidth,angle=0]{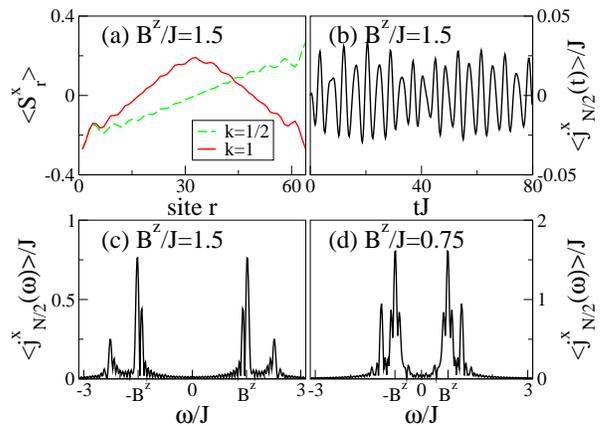}
\caption{(color online) Simulation of the transverse spin-current dynamics
using tDMRG: (a) Initial magnetization profiles $\langle S^x_r \rangle$ at $t=0$ for
different $k$; (b) Coherent oscillation of the spatially averaged current $\langle
j^x_{N/2}(t) \rangle$ for $k=0.5$ and $B^z/J = 1.5$; (c), (d) Discrete Fourier
transform for (b) and for $B^z/J = 0.75$. Besides the dominant peak at the Larmor
frequency $\omega_L = B^z$, there is another significant peak at a higher frequency
$\omega_L + \delta \omega$. For more details, see Ref.~\onlinecite{epaps}.}
\label{fig:RTcurrents}
\end{figure}

Now we turn to the real-time evolution of the spin-current derived from a
Krylov-space based tDMRG approach \cite{tDMRG}. This allows us to study larger
systems than with ED, however, at zero temperature and only below the
saturation field. The latter follows since there are no current-carrying states
for $B^z > B^z_c$ at $T=0$. Moreover, since ED and AAA already suggest that
$\tau \rightarrow \infty$ as $T \rightarrow 0$, limitations in the accessible
simulation times
confine the tDMRG to an analysis of $\delta
\omega$. To induce a
current we add a perturbation $H_1 = \sum_r B^x_r
S^x_r$ to Eq.~(\ref{H}) with $B^x_r = B^x \cos(2\pi k r/N)$. First, the ground
state of $H + H_1$ is evaluated using DMRG, then the system is left to evolve
under $H$ alone.

Typical transverse magnetization profiles $\langle S^x_r \rangle$ at $t=0$ are
shown in Fig.~\ref{fig:RTcurrents}~(a) for $(B^z,B^x)/J = (1.5,1)$ and
for small values of $k$. $\langle S^x_r \rangle$ follows $B^x_r$ qualitatively, with
additional $2k_F$-oscillations from the underlying Luttinger liquid. We perform
the time evolution using $m=200$ states for the ground-state calculation, a time
step of $\delta t J = 0.25$, and a fixed discarded weight \cite{tDMRG, epaps}. Although $H_1$
breaks $U(1)$ symmetry, we can still obtain reliable results for  
$L \leq 64$ lattice sites.
To analyze currents free of spatial oscillations we
coarse-grain the data by averaging over suitable parts of the chain. Figure
\ref{fig:RTcurrents}~(b) shows an example of the time evolution of the
current $\langle j_{N/2}^x(t) \rangle$, averaged over the left half of the chain, at
$(B^z,B^x)/J = (1.5,1)$ for $k=0.5$.
For the times reached in the simulation ($t J \approx 80$), no relaxation can be
observed. This is consistent with the expectation for
$T = 0$ drawn from ED and AAA. In view of
the substantial number of periods covered in Fig.~\ref{fig:RTcurrents}~(b) we
have chosen to directly study the discrete Fourier transform of the real-time
data in order to obtain the \emph{dominant} frequencies. This is shown in
Figs.~\ref{fig:RTcurrents}~(c) and (d) for $\langle j_{N/2}^x(t) \rangle$ at
$(B^z,B^x)/J =(1.5,1)$ and $(B^z,B^x)/J = (0.75,1)$ at $k=0.5$. These two figures
clearly evidence the main result from tDMRG, namely that, fully consistent
with the findings from ED and AAA, there are two characteristic frequencies in
the current dynamics, namely $\omega_L$ and $\omega_L + \delta \omega$. In
contrast to the linear-response regime, the analysis of the relaxation
scenario finds
the contribution at $\omega_L$ to be the larger one. This is not surprising
since two different scenarios
are compared, characterized by similar yet different correlation functions. We have
checked that the results of Fig.~\ref{fig:RTcurrents}~(b)-(d) are insensitive (i) to
the details of the coarse-graining, (ii) to varying $k$ within the small $k$-regime,
and (iii) to the strength of $B^x$, at least up to $B^x/J = 1$, as used here
\cite{epaps}.
The latter implies a minor role of non-linearity (non-equilibrium).

\begin{figure}[tb]
\includegraphics[width=0.6\linewidth]{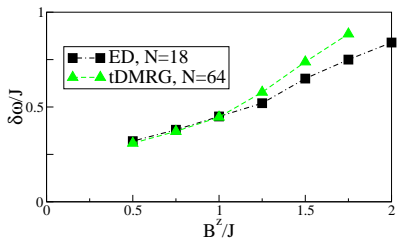}
\caption{(color online) Frequency shift $\delta \omega$ w.r.t.~the
magnetic field $B^z$, as obtained from ED ($\beta J = 3$) and tDMRG.}
\label{figure_frequency_shift}
\end{figure}

Finally, in Fig.~\ref{figure_frequency_shift} we compare $\delta\omega$ from ED
with tDMRG for $0.5\leq B^z/J \leq 2$. The agreement is remarkably good indeed, not
only in view of the 
different scenarios. For intermediate
fields ($B^z/J = 0.5$, $0.75$, $1$) the frequency shifts
match each other almost exactly, while we attribute the slight deviation
of ED from tDMRG
at larger fields to finite temperature effects, where convergence
to the $T=0$ values has not been reached yet [see $B^z/J = 3$ in
Fig.~\ref{figure_frequency_shift_decay_time}~(b)]. For $B^z/J < 0.5$, the
accessible time scales prevent a reliable determination of $\delta \omega$ in
our approaches.

In summary, we studied the transverse spin-current dynamics in the spin-1/2
Heisenberg chain. As a main result, besides a coherent oscillation at the Larmor
frequency, we provided evidence for a second nontrivial collective oscillation
at higher frequencies, emerging at low temperatures as a genuine many-magnon effect 
and turning coherent as the temperature is lowered.
\begin{acknowledgments}
{\it Acknowledgments}
This work was supported by the {\it Deutsche Forschungsgemeinschaft} through FOR
912.
\end{acknowledgments}

\section{Supplemental Material}

\subsection{I.~Derivation of the Asymptotic Approximation}

In this section we give a detailed derivation of the asymptotic approximation in
Eq.~(4) of the Letter. For convenience, we shift the zero point of the energy
$E$ to $N (J/4 - B^z/2)$ and of the quantum number $M$ to $N/2$. The shifted
quantities will be denoted by $\cal E$ and $\cal M$ in the following,
\begin{equation}
{\cal E} = E - N(\frac{J}{4} - \frac{B^z}{2}) \, , \quad {\cal M} = -(M -
\frac{N}{2}) \, .
\end{equation}

\begin{figure}[t]
\includegraphics[width=1.0\linewidth]{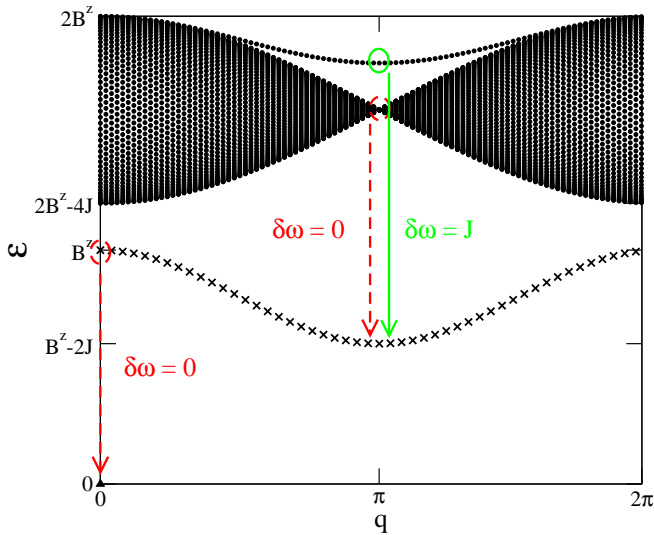}
\caption{The spectrum of the Hamiltonian in the ${\cal M}=0$-, $1$-, and
$2$-magnon subspaces (symbols). Additionally, those transitions of the transverse
current are indicated that are relevant for the asymptotic approximation at low
temperatures above the critical magnetic field: `Forbidden' transitions
(red, dashed lines) yielding no contribution as well as `allowed' transitions
(green, solid lines) leading to the dominant contribution. The latter transitions
involve a two-magnon band at the edge of the Brillouin zone, i.e., $q = \pi$.
The associated frequencies are larger than the Larmor frequency $\omega_L = B^z$,
namely, these frequencies are shifted by $\delta \omega = J$.}
\label{figure_spectrum}
\end{figure}

Above the critical field the ground state is fully polarized
\begin{equation}
\psi_{({\cal M}=0, q=0)} = | \! \uparrow \uparrow \ldots \uparrow
\uparrow \rangle
\end{equation}
with momentum $q=0$ and energy ${\cal E}_{(0,0)} = 0$, see
Fig.~\ref{figure_spectrum} (triangle). Periodic boundary conditions are assumed
in this section. This state is an eigenstate of the current operator with the
eigenvalue zero
\begin{equation}
j^x \, \psi_{(0,0)} = 0 \, .
\end{equation}
In the limit of $T=0$ only matrix elements from this initial state contribute to
the current autocorrelation function $C^x(\omega)$. Therefore
$\lim_{T\rightarrow 0}C^x(\omega)=0$. At any finite temperature, $\beta J\gg 1$,
initial states from the subspace with ${\cal M} = 1$, i.e.~one-magnon states,
start to contribute to the current autocorrelation function. Using the operator
$T^\mu$, which translates a state by $\mu$ sites, the eigenstates of the
Hamiltonian $H$ in Eq.~(1) of the Letter read in this subspace
\begin{equation}
\psi_{(1,q)} = \frac{1}{\sqrt{N}} \sum_{\mu = 0}^{N-1} e^{\imath q \mu}
\, T^\mu \, | \! \downarrow \uparrow \uparrow \ldots \uparrow
\uparrow \rangle \, ,
\label{1pts}
\end{equation}
$q = 2 \pi k/N$, $k = 0, 1, \ldots, N-1$. Their energies are
\begin{equation}
{\cal E}_{(1,q)} = J \, (\cos q - 1) + B^z \, ,
\label{sup1}
\end{equation}
see Fig.~\ref{figure_spectrum} (crosses). At non-zero momentum these one-magnon
states obviously carry a net-current, however, they are not eigenstates of the
current operator. In fact one readily obtains
\begin{equation}
j^x \, \psi_{(1,q)} = \imath \, \frac{J}{2} \, (1- e^{\imath
q}) \, \varphi_{(2,q)} \, , \label{eq_transition}
\end{equation}
where $\varphi_{(2,q)}$ are states from the subspace with ${\cal M} = 2$,
i.e.~the two-magnon subspace. They read
\begin{equation}
\varphi_{(2,q)} = \frac{1}{\sqrt{N}} \sum_{\mu = 0}^{N-1} e^{\imath
q \mu} \, T^\mu \, | \! \downarrow \downarrow \uparrow
\uparrow \ldots \uparrow \uparrow \rangle \, .
\end{equation}
In general, and in contrast to Eqs.~(\ref{1pts}), (\ref{sup1}), these states are
{\em no} eigenstates of the Hamiltonian $H$ in Eq.~(1) of the Letter, but
\begin{equation}
H \, \varphi_{(2,q)} = (2 B^z - J) \, \varphi_{(q,2)} + \frac{J}{2}
(1 + e^{\imath q}) \, \tilde{\varphi}_{(2,q)} \, ,
\label{2ptsg}
\end{equation}
where $\tilde{\varphi}_{(2,q)}$ refers to states generated by the transverse
fluctuations of $H$, which separate the two adjacent flipped spins by one
site. However, in the vicinity of $q = \pi$, the prefactor $(1 + e^{\imath q})$
suppresses these contributions rendering $\varphi_{(2,\pi)}$ an exact eigenstate
of $H$. From its eigenenergy $(2 B^z - J)$ and Fig.~\ref{figure_spectrum} (green
circle), it is obvious that this state is the (anti)bound two-magnon state
$\psi^r_{(2,q)}$ at $q=\pi$, well known from Bethe-Ansatz. Its dispersion over
the complete Brillouin zone is \cite{Bethe1931}
\begin{equation}
{\cal E}^r_{(2,q)} = \frac{J}{2} \, (\cos q -1) + 2 B^z
\end{equation}
and is situated above the two-magnon continuum.

Due to their Boltzmann weight the transitions from $\psi_{(1,q)}$ into
$\varphi_{(2,\pi)}$ at $q\approx\pi$ dominate the current autocorrelation function
asymptotically for $\beta J \gg 1$. Therefore, the leading contribution to $C^x$
results from projecting all intermediate states {\em solely} onto $\psi^r_{(2,q)}$.
Since the antibound state is separated by a gap of ${\cal O}(J)$ from the two-magnon
continuum at $q\approx\pi$, we may use $\langle \varphi_{(2,q)} | \, \psi^r_{(2,q)}
\rangle \approx 1$ in that region, leading to
\begin{eqnarray}
\tilde{C}^x_\text{AAA}(\omega) \! &\approx& \! \frac{1}{N} \sum_q e^{-\beta
{\cal E}_{(1,q)}} \, | \langle \psi_{(1,q)} | j^x \, \psi^r_{(2,q)}
\rangle |^2
\, \nonumber\\
&& \! \times \, \delta(\omega - [{\cal E}^r_{(2,q)} - {\cal
E}_{(1,q)}]) \, .
\end{eqnarray}
Using Eq.~(\ref{eq_transition}) and introducing the frequency $\omega_q = J/2 \,
(1 - \cos q)$ this can be rewritten as
\begin{equation}
\tilde{C}^x_\text{AAA}(\omega) \approx \frac{J}{N} \, e^{-\beta B^z} \,
\sum_q \omega_q \, e^{2 \beta \omega_q} \, \delta(\omega - [B^z + \omega_q])
\, .
\end{equation}
Since momentum enters only through $\omega_q$, we may introduce the
corresponding density of states and replace the sum by an integral with respect
to $\omega_q$. This results in
\begin{equation}
\tilde{C}^x_\text{AAA}(\omega + B^z) \approx \frac{J}{2 \pi} \,
 \, \frac{e^{\beta (2\omega-B^z)}}{\sqrt{J/\omega - 1}} \, \Theta(\omega)
\, \Theta(J - \omega) \, ,
\end{equation}
where $\Theta(\omega)$ is the Heavyside function. Fourier transforming this
to the time domain we get
\begin{eqnarray}
\tilde{C}^x_\text{AAA}(t) \!\! &\approx& \!\! \frac{J^2}{2} \, e^{-\beta(B^z
- J)} \, e^{\imath(B^z + J/2)
t} \nonumber \\
&& \!\! \times \left [ {\cal I}_0(J \, [\beta + \frac{\imath \,
t}{2}]) + {\cal I}_1(J \, [\beta + \frac{\imath \, t}{2}]) \right ] \, ,
\end{eqnarray}
where ${\cal I}_{0,1}(z)$ are modified Bessel functions of the first kind. At
low temperatures, i.e.~for $\beta J\gg 1$, we may insert their asymptotic
forms for $|z|\gg 1$, which are ${\cal I}_{0,1}(z) \approx e^z/\sqrt{2 \pi \,
z}$. Therefore
\begin{equation}
\tilde{C}^x_\text{AAA}(t) \approx \sqrt{\frac{J^3}{\pi}} \, e^{-\beta(B^z
- 2 J)} \frac{e^{\imath (B^z + J) t}}{\sqrt{2 \beta + \imath \, t}} \, .
\label{asyfin}
\end{equation}
The real part of this, i.e.~$C^x_\text{AAA}(t) = \text{Re} \,
\tilde{C}^x_\text{AAA}(t)$, is Eq.~(4) of our Letter, with $\omega_L=B^z$.

\begin{figure}[t]
\includegraphics[width=0.8\linewidth]{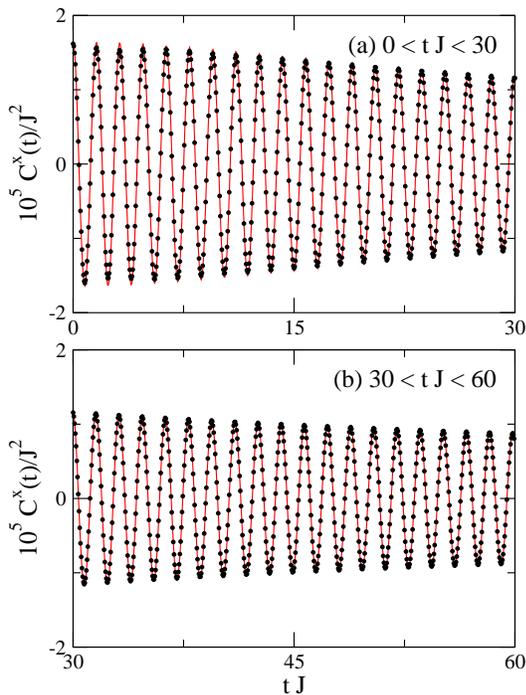}
\caption{The current autocorrelation function $C^x(t)$ for (a) $0 \leq t J \leq
30$  and (b) $30 \leq t J \leq 60$, restricted to the transitions into the subspaces
${\cal M} \leq 1$ (one-magnon subspace). Data are evaluated numerically by the use
of ED (circles) and are shown for the parameters $\beta J = 9$, $B^z/J = 3$, and
$N=200$. For comparison, the asymptotic approximation according to Eq.~(4) of the
Letter is included (curves).}
\label{figure_formula}
\end{figure}

To assess the quality of the asymptotic approximation, we compare this result
with ED for the current autocorrelation function above the saturation field,
restricting the intermediate-state Hilbert space to two-particle states. This
corresponds to taking the limit $\beta J\gg 1$. Due to this restriction ED is
possible for rather large systems sizes $N$. We choose $N = 200$. As shown in
Fig.~\ref{figure_formula}, the agreement is excellent. This validates the
approximations involved in going from Eq.~(\ref{eq_transition}) to
(\ref{asyfin}).

\subsection{II.~ED Calculations}

\begin{figure}[t]
\includegraphics[width=0.8\linewidth]{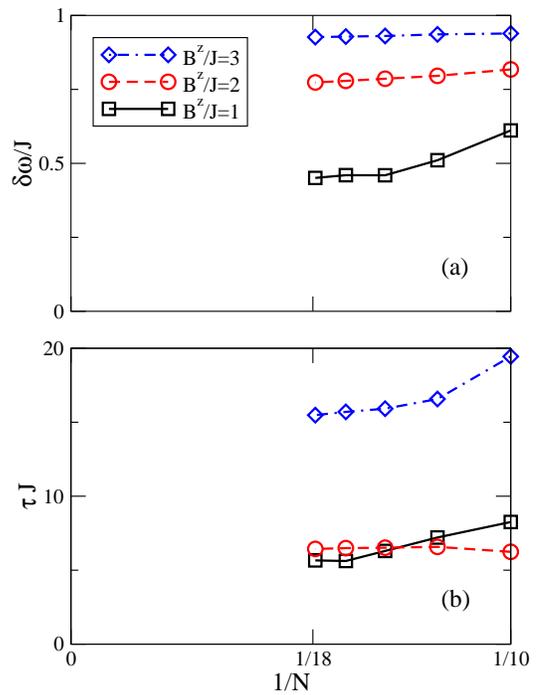}
\caption{The extracted (a) frequency shift $\delta \omega$ and (b) decay time
$\tau$ as a function of the inverse size $1/N$ for the parameters $\beta J = 2$
and $B^z/J = 1$, $2$, and $3$ (symbols). Lines are guides to the eyes.}
\label{figure_finitesize}
\end{figure}

In this section we provide supplementary material on
our ED calculations concerning a potential impact of the finite system
size. Specifically, we  demonstrate that for the systems of size $N=18$ and
time scales $tJ\lesssim 20$, as used in the Letter, finite-size effects in
the frequency shift $\delta\omega$ and the relaxation time $\tau$ can be
neglected. {\em Both} of these quantities are determined by assuming a single
exponential $R(t) = \exp(-t / \tau)$ to model the envelope of the decay of the
coherent oscillation as in Eq.~(3) of the Letter. This procedure
leads to satisfactory fits to the ED results for not too low temperatures,
$\beta J\lesssim 3$. Examples of this are shown in Fig.~2 of the
Letter.

Figure \ref{figure_finitesize} summarizes our results for $\delta\omega$ and
$\tau$ as obtained from $N=10$, $12$, $14$, $16$, and $18$ for various magnetic
fields at $\beta J=2$. This figure clearly demonstrates that both, $\delta\omega$
and $\tau$ either display almost no finite size dependence or a clear tendency
towards saturation as a function of $1/N$. In all cases shown, the absolute changes
in going from $N=16$ to $18$ are negligible. Significant finite size effects can
only be seen at systems sizes $N\leq 14$. Therefore, for the temperatures and fields
considered in the Letter, it is justified to use $N=18$ data to obtain $\delta\omega$
and $\tau$ (see Fig.~3 of the Letter).

\subsection{III.~DMRG Calculations}

In this section we provide details of our DMRG simulations of the real-time
evolution of the transverse current. We will focus on three aspects: (i) the
coarse graining of the current, (ii) the dependence on the initial state and its
characteristic wave length $k$, and (iii) the numerical determination of the
frequency shift $\delta \omega$, depending on the maximum simulation time and
system size. The simulations were carried out with a fixed discarded weight,
which we have varied from  $10^{-4}$ to $10^{-6}$ in order to check convergence
of our results.

\begin{figure}[t]
\center
\includegraphics[width=0.9\linewidth]{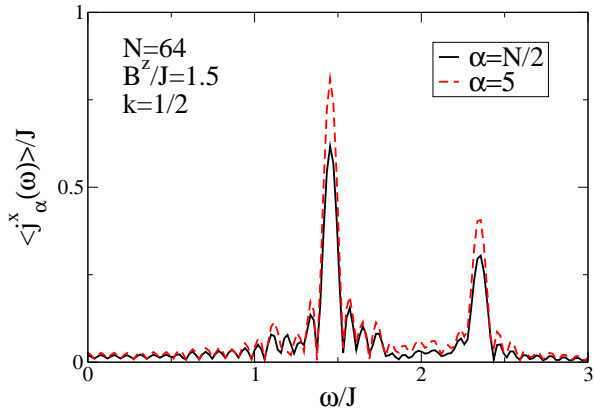}
\caption{Coarse graining: The solid, black line is the frequency-dependent
current averaged over the left half of the system while the dashed, red line is
averaged over five sites, counting away form the center of the system with $N=64$
(to be precise, these are sites $ r=30$, $31$, $32$, $33$, $34$). The main features,
i.e., the dominant frequencies do not depend on the coarse graining.}
\label{fig:coarse}
\end{figure}

Figure~\ref{fig:coarse} shows the Fourier spectrum of the transverse current
for two different coarse graining schemes. The solid, black line is taken from
Fig.~4(c) of the Letter where we average the current over the left half of the
chain (labeled by $\alpha=N/2$ in Fig.~\ref{fig:coarse}). The dashed, red line is
the result from averaging the current over five sites in the middle of the
chain (denoted by $\alpha=5$); in this example, the coarse graining is taken over
sites $r=30$, $31$, $32$, $33$, $34$ in a system with $N=64$. As the figure clearly
shows, the position of the two maxima, i.e., the one at the Larmor frequency
$\omega_L$ and the one at $\omega_L+\delta\omega$, as well as their (relative)
weights are insensitive to the coarse graining.

\begin{figure}[b]
\center
\includegraphics[width=0.9\linewidth]{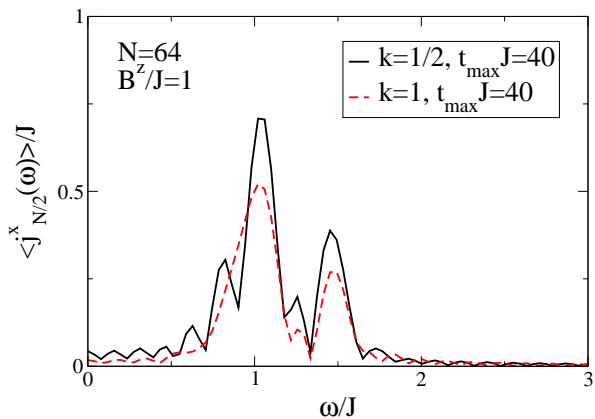}
\caption{Dependence on the initial state: Spectrum of the transverse current for
two different values of $k$: $k=1/2$ (solid, black line) and $k=1$ (dashed, red
line). In both cases, $B^x/J=1$, $B^z/J=1$ and $N=64$. The Fourier transform is
taken with $t_{\mathrm{max}}J=40$ due to the higher numerical costs at larger
$k$.}
\label{fig:kdep}
\end{figure}

Turning to the initial states of the real-time evolution, they are constructed
by adding a site-dependent transverse field via a term $\sum_r B^x_r S^x_r$ to
the Hamiltonian with $B^x_r=B^x\cos(2\pi kr/N)$. Figure~\ref{fig:kdep}
illustrates that a central result of our Letter, namely the positions of the
maxima in the current's Fourier spectrum, does not depend on the wave vector $k$
of the perturbing field $B_r^x$ in the long wave-length limit. Computational
constraints lead to a decrease of the maximum simulation times accessible as $k$
is increased. For a comparison of $k=1/2$ and 1 this implies that we have to
confine ourselves to $t J\lesssim 40$. Therefore, the features in
Fig.~\ref{fig:kdep} are broader than in Fig.~4~(c) of our Letter.

\begin{figure}[t]
\center
\includegraphics[width=1.0\linewidth]{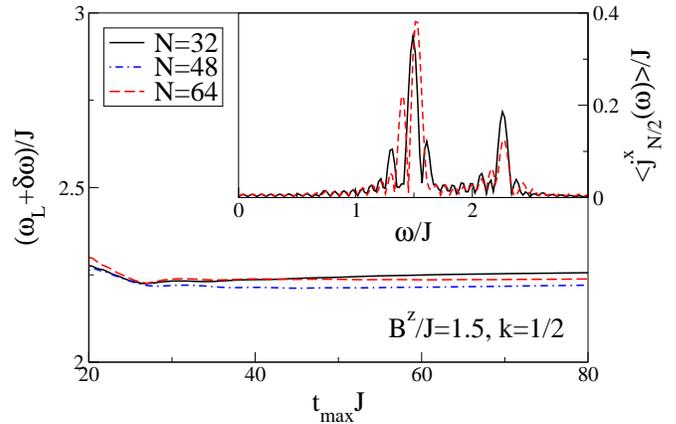}
\caption{Numerical determination of $\omega_L+\delta \omega$ in dependence of
the simulation time, for three different system sizes: $N=32$ (solid, black
line), $N=48$ (dash-dotted, blue line) and $N=64$ (dash-dotted, red line). The
final values are all within the overall numerical accuracy and no systematic
finite size effects are visible. The inset shows the spectra at $t_\mathrm{max} J
=80$ for $N=32$ (solid, black line) and $N=64$ (dashed, red line).}
\label{fig:tmax}
\end{figure}

In Fig.~\ref{fig:tmax}, we address the convergence of the position of the second
maximum in $j^x_{N/2}(\omega)$ at $\omega_L+\delta\omega$ as a function of the
simulation time and the system size. Since we perform a discrete Fourier
transform on a finite time window to obtain $j^x_{N/2}(\omega)$ from the
real-time data $j^x_{N/2}(t)$, the resulting spectrum depends on the maximum
simulation time. After a minimum time needed to resolve the two frequencies has
been reached we compute the spectrum after each time step and extract both
frequencies. The quantity displayed in the figure is $\omega_L+\delta\omega$
averaged over these spectra obtained from $t_{\mathrm{max}} J=20$ up to the
maximum simulation time $t_{\mathrm{max}}$. For $N=32$ and $48$ (solid, black and
dash-dotted, blue line, respectively), $\omega_L+\delta \omega$ weakly increases
with $t_{\mathrm{max}}$, while for $N=64$ (dashed red line), the convergence is
much faster. The finite-size effects in $\omega_L+\delta\omega$ are
non-monotonous and result in a small uncertainty of about 2-3$\%$, well within
the overall numerical accuracy of the simulations.  The inset shows the spectrum
of the current as a function of frequency at $t_\mathrm{max} J=80$ for $N=32$ and
$N=64$. While the dominant frequencies are only slightly affected by finite-size
effects, spurious additional peaks appear on the smaller system that are
irrelevant for the results and discussion presented in the Letter.

\subsection{IV.~Comment on Units}

In the Letter all quantities are expressed in units of the exchange coupling
constant $J$. Moreover, and to abbreviate the theoretical analysis, Planck's constant
$\hbar$ has been set to unity as usual. Similarly, the Bohr magneton
$\mu_\mathrm{B}$ and the spin Land\'{e} factor $g_\mathrm{S}$ have been
absorbed into the definition of the magnetic field. The correspondence
between these units and SI-units is summarized in Tab.~\ref{units}.
Therein, $J$ is expressed in units of temperature, i.e.,
divided by the Boltzmann constant $k_\mathrm{B}$.

\begin{table}[h]
\begin{tabular}{|l|l|l|}
\hline
\multicolumn{1}{|c|}{\multirow{2}{*}{\bf quantity}} & \multicolumn{2}{|c|}{\bf unit} \\
\cline{2-3}
& \multicolumn{1}{|c|}{\bf Letter} & \multicolumn{1}{|c|}{\bf SI} \\
\hline
magnetic field & $J$   & $k_\mathrm{B}/(\mu_\mathrm{B} \, g_\mathrm{S}) \,\, (J/k_\mathrm{B})$ \\
               &       & $\approx 0.744 \, \mathrm{T}/\mathrm{K} \,\, (J/k_\mathrm{B})$ \\
\hline
frequency      & $J$   & $k_\mathrm{B}/\hbar \,\, (J/k_\mathrm{B})$ \\
               &       & $\approx 1.309 \cdot 10^{11} \, \mathrm{Hz}/\mathrm{K} \,\, (J/k_\mathrm{B})$ \\
\hline
time           & $1/J$ & $\hbar/k_\mathrm{B} \,\, (k_\mathrm{B}/J)$ \\
               &       & $\approx 7.638 \cdot 10^{-12} \, \mathrm{s} \mathrm{K} \,\, (k_\mathrm{B}/J)$ \\
\hline
\end{tabular}
\caption{Correspondence between the units in the Letter and SI-units.}
\label{units}
\end{table}

\end{document}